\documentclass[a4paper,11pt]{article}
\pdfoutput=1 \usepackage{jheppub} \usepackage[T1]{fontenc} \RequirePackage[displaymath]{lineno}

\usepackage{epsfig}
\usepackage{graphicx}\usepackage{dcolumn}\usepackage{bm}\usepackage{ltablex,booktabs}
\usepackage{overpic}
\usepackage{subfigure}
\usepackage{float}
\usepackage{color}
\usepackage{amsmath}
\usepackage{mathcomp}
\usepackage{mathrsfs}
\usepackage{multirow}
\usepackage{rotating}
\usepackage{amssymb}
\usepackage{gensymb}
\usepackage{amsmath}
\usepackage{tabularx}
\usepackage{epstopdf}
\usepackage{ulem}
\usepackage{xcolor}

\title{Observation of \boldmath $\psi(3686) \to K^{-}\Lambda(1520)\bar{\Xi}^{+} + c.c.$}
\collaborationImg{\includegraphics[width=0.15\textwidth, angle=90]{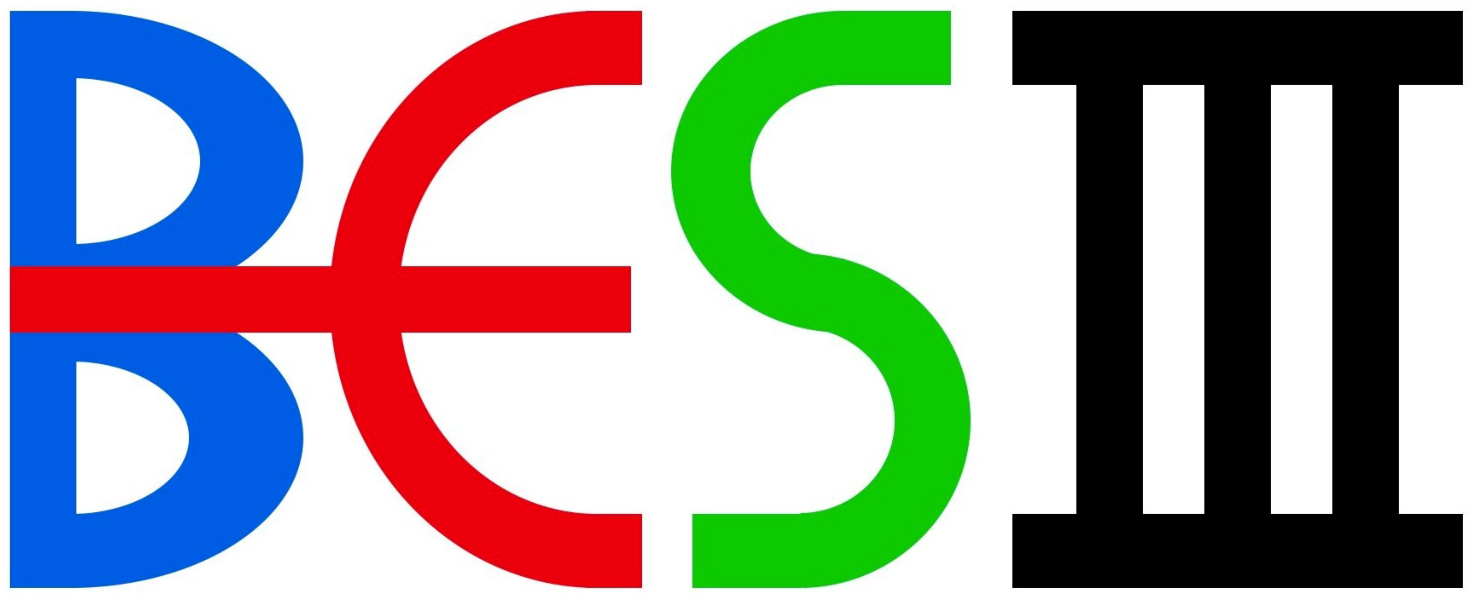}}
\collaboration{The BESIII Collaboration}

\date{\today}
\abstract{Based on $(2712.4 \pm 14.3)\times 10^6$ $\psi(3686)$ events collected at the BESIII detector operating at the BEPCII collider, we present the first observation of the decay $\psi(3686) \to K^{-}\Lambda(1520)\bar{\Xi}^{+} + c.c.$. The product branching fraction ${\cal B}[\psi(3686) \to K^{-}\Lambda(1520)\bar{\Xi}^{+} + c.c.] \times {\cal B}[\Lambda(1520) \to pK^{-}]$ is measured to be $(9.5 \pm 0.8 \pm 1.1) \times 10^{-7}$, where the first uncertainty is statistical and the second systematic.}

\keywords{Charmonium Physics, Three-Body Baryonic Decay, Branching Fraction, $e^{+}e^{-}$ Collision}
\arxivnumber{}
\begin{document}
\maketitle
\flushbottom

\section{Introduction}
The discovery of $J/\psi$ and other charmonium states of $c\bar{c}$ played important roles in the development of the theory of the strong interaction in the Standard Model (SM)~\cite{1974sol,1974ind}. These states can probe a wide range of energy scales in the Quantum Chromodynamics (QCD) from high-energy to low-energy regions, where the non-perturbative effects become dominant~\cite{Asner:2008nq}. Experimental study on hadronic decays of charmonium states is a bridge between perturbative QCD and non-perturbative QCD. Many topics involving strong interaction can be investigated, such as color octet and singlet contributions, the violation of helicity conservation, and SU(3) flavor symmetry breaking effects~\cite{Asner:2008nq,Rybicki:2009zza}. Since baryons represent the simplest system in which three colors of quarks neutralize into colorless objects with the essential non-Abelian character of QCD, a systematic
study of baryon spectroscopy can provide critical insights into the nature of QCD in the confinement domain.

Compared to two-body final states, the theoretical calculation for three-body decays of charmonium states is more challenging. The decays of charmonium states into three-body final states are helpful in the search for excited baryon states and threshold enhancements, and numerous such decay modes have been studied recently~\cite{BESIII:2024zav}. However, the study of baryon spectroscopy remains incomplete, with many of the states predicted by SU(3) multiplets yet to be discovered or well-established. Most published measurements of $\psi(3686)$ decays involve zero or one strange quark~\cite{ParticleDataGroup:2024cfk}. The knowledge of excited baryon states with two strange quarks, $i.e.$ $\Xi^{*}$ hyperons, is particularly limited due to their small production rates and complicated decay topology. To date, only several states have been observed, and few of them have well-determined spin and parity. Further searches and detailed investigations into excited baryons are important to further understand the QCD mechanism.

In this paper, we report the first observation of the decay $\psi(3686) \to K^{-}\Lambda(1520)\bar{\Xi}^{+} + c.c.$, and the measurement of the product branching fraction ${\cal B}[\psi(3686) \to K^{-}\Lambda(1520)\bar{\Xi}^{+} + c.c.] \times {\cal B}[\Lambda(1520) \to pK^{-}]$, based on $(2712.4\pm 14.3)\times 10^6$ $\psi(3686)$ events collected with the BESIII detector~\cite{BESIII:2024lks}. Throughout this paper, the charge conjugation decay mode is always implied.

\section{BESIII detector and Monte Carlo simulation}
\label{sec:detector_dataset}

The BESIII detector is a magnetic spectrometer~\cite{ABLIKIM2010345,Ablikim_2020} located at the Beijing Electron Positron Collider (BEPCII)~\cite{Yu:IPAC2016-TUYA01}, which operates with a peak luminosity of $1.1 \times 10^{33} \rm{cm}^{-2} \rm{s}^{-1}$ in the center-of-mass (CM) energy range from 1.85 to 4.95 GeV. A helium-based multilayer drift chamber (MDC), a plastic scintillator time-of-flight system (TOF), and a CsI(Tl) electromagnetic calorimeter (EMC) compose the cylindrical core of the BESIII detector, and they are all enclosed in a superconducting solenoidal magnet providing a 1.0 T magnetic field. The solenoid is supported by an octagonal flux-return yoke with resistive plate counter muon identifier modules interleaved with steel. The acceptance of charged particles and photons is 93\% over a 4$\pi$ solid angle. The charged-particle momenta resolution at 1.0 GeV/$c$ is 0.5\%, and the specific energy loss ($dE/dx$) resolution is 6\% for the electrons from Bhabha scattering at 1 GeV. The EMC measures photon energies with a resolution of 2.5\%(5\%) at 1 GeV in the barrel (end-cap) region. The time resolution of the TOF barrel part is 68 ps, while that of the end-cap part is 110 ps. The end-cap TOF was upgraded in 2015 with multi-gap resistive plate chamber technology, providing a time resolution of 60 ps~\cite{etof1,etof2,etof3}, which benefits 85$\%$ of the data used in this analysis.

Simulated samples produced with a {\sc geant4}-based~\cite{GEANT4:2002zbu} Monte Carlo (MC) package, which includes the geometric description of the BESIII detector and the detector response, are used to determine detection efficiency and estimate backgrounds. The simulation models the beam energy spread and initial state radiation in the $e^{+}e^{-}$ annihilations with the generator {\sc kkmc}~\cite{Jadach:2000ir, Jadach:1999vf}. The inclusive MC sample includes the production of the $\psi(3686)$ resonance, the initial-state radiation production of the $J/\psi$ meson, and the continuum processes incorporated in {\sc kkmc}~\cite{Jadach:2000ir, Jadach:1999vf}. All particle decays are modelled with {\sc evt}{\sc gen}~\cite{Lange:2001uf, EVTGEN2} using branching fractions either taken from the Particle Data Group (PDG)~\cite{ParticleDataGroup:2024cfk}, when available, or otherwise estimated with {\sc lundcharm}~\cite{Chen:2000tv, LUNDCHARM2} for the unknow ones. Final state radiation from charged final state particles is incorporated using {\sc photos}~\cite{PHOTOS}. To determine the detection efficiency, a signal MC sample of $\psi(3686) \to K^{-}\Lambda(1520)\bar{\Xi}^{+}, \bar{\Xi}^{+} \to \bar{\Lambda}\pi^{+}, \bar{\Lambda} \to \bar{p}\pi^{+}$, and $\Lambda(1520) \to pK^{-}$ is generated uniformly in phase-space (PHSP). An inclusive $\psi(3686)$ MC sample, consisting of $2747 \times 10^{6}$ events, is used to estimate potential backgrounds. The data sample collected at a CM energy of 3.773 GeV, corresponding to an integrated luminosity of 2.93 fb$^{-1}$, is used to estimate the continuum background.

\section{Event selection}\label{EVT-selection}
The cascade decay of interest is $\psi(3686) \to K^{-}\Lambda(1520)\bar{\Xi}^{+}$, $\Lambda(1520) \to pK^{-}$, where $\bar{\Xi}^{+} \to \bar{\Lambda}\pi^{+}$, and $\bar{\Lambda}\to \bar{p}\pi^{+}$. In track-level selection, at least six charged tracks are reconstructed within the polar angle ($\theta$) range of $|\cos\theta|<0.93$, where $\theta$ is defined with respect to the $z$-axis, which is the symmetry axis of the MDC. Since the bachelor kaon and the two tracks from $\Lambda(1520)$ decay have no secondary vertex, we further require at least three tracks to originate from the interaction point (IP), i.e. ${V_{r}} < 1$ cm, $\left| {{V_z}} \right| < 10$ cm, where $V_r$ and $V_z$ are the closest approaches of the tracks to the IP in transverse plane and in $z$ coordinate respectively. For each charged track, particle identification (PID) is performed. At least six charged particles, $p\bar{p}K^{-}K^{-}\pi^{+}\pi^{+}$, are identified by combining measurements of the $dE/dx$ in the MDC and the flight time in the TOF to form PID likelihoods for each particle hypothesis. If there are multiple combinations of $p\bar{p}K^{-}K^{-}\pi^{+}\pi^{+}$, the combination with the largest sum of PID likelihoods is kept for further analysis. The $\bar{\Xi}^+$ is tagged by the recoil mass of proton and two kaons, $RM(pK^{-}_{1}K^{-}_{2})$. Here (and elsewhere), $K^{-}_{1}$ denotes the kaon from $\Lambda(1520)$ decay, while $K^{-}_{2}$ denotes the bachelor. The $K_1$ is assigned by requiring its invariant mass $M(pK_1^{-})$ to be closer to the mass of $\Lambda(1520)$.

\begin{figure}[htp]
\begin{center}
\begin{minipage}[t]{0.6\linewidth}
\includegraphics[width=1\textwidth]{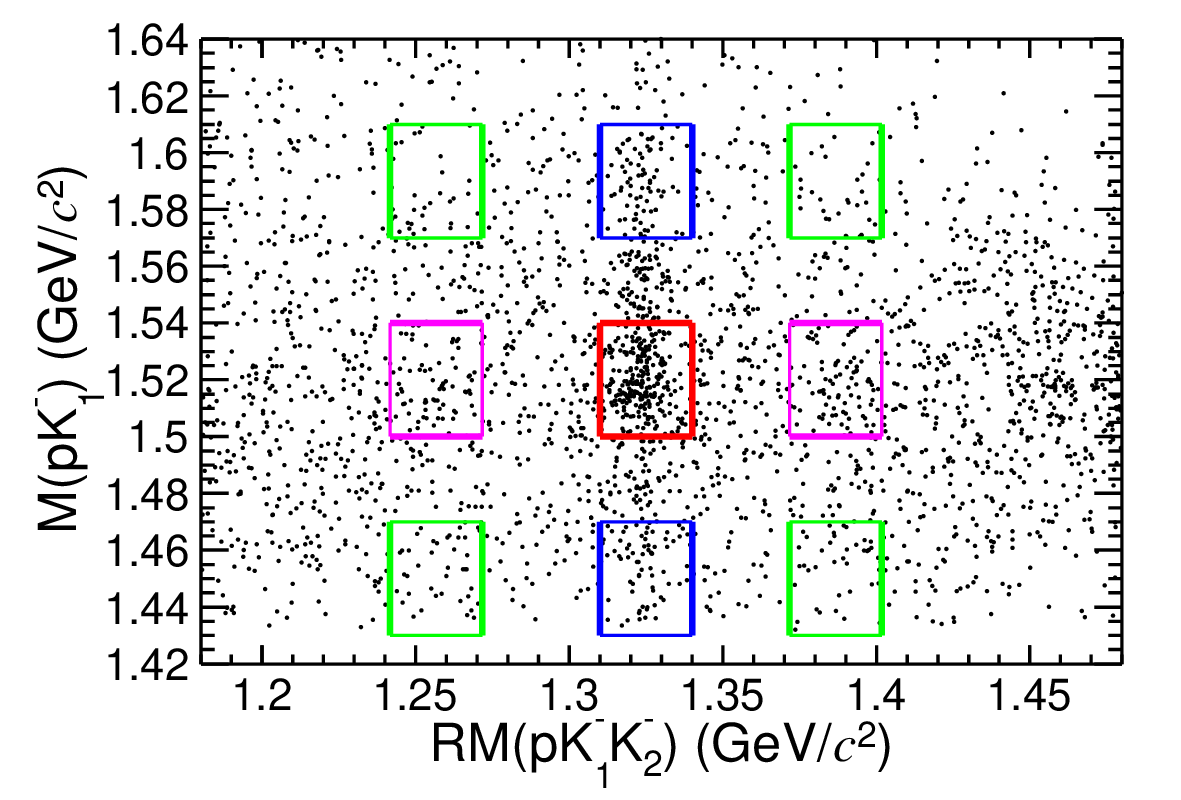}
\end{minipage}
\caption{The distributions of $RM(pK^{-}_{1}K^{-}_{2})$ versus $M(pK^{-}_{1})$ for data events.}
\label{fig:2dsideband}
\end{center}
\end{figure}

\begin{figure}[htp]
\begin{center}
\begin{minipage}[t]{0.49\linewidth}
\includegraphics[width=1\textwidth]{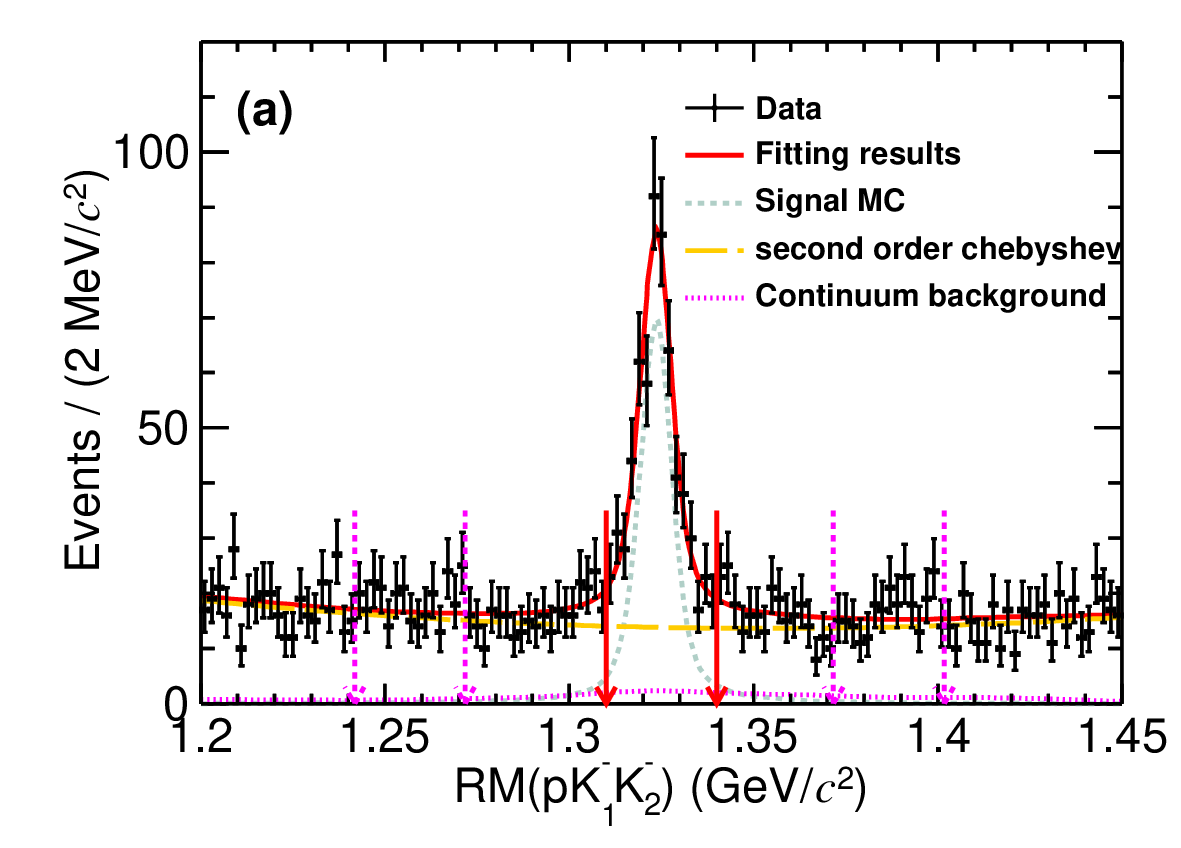}
\end{minipage}
\begin{minipage}[t]{0.49\linewidth}
\includegraphics[width=1\textwidth]{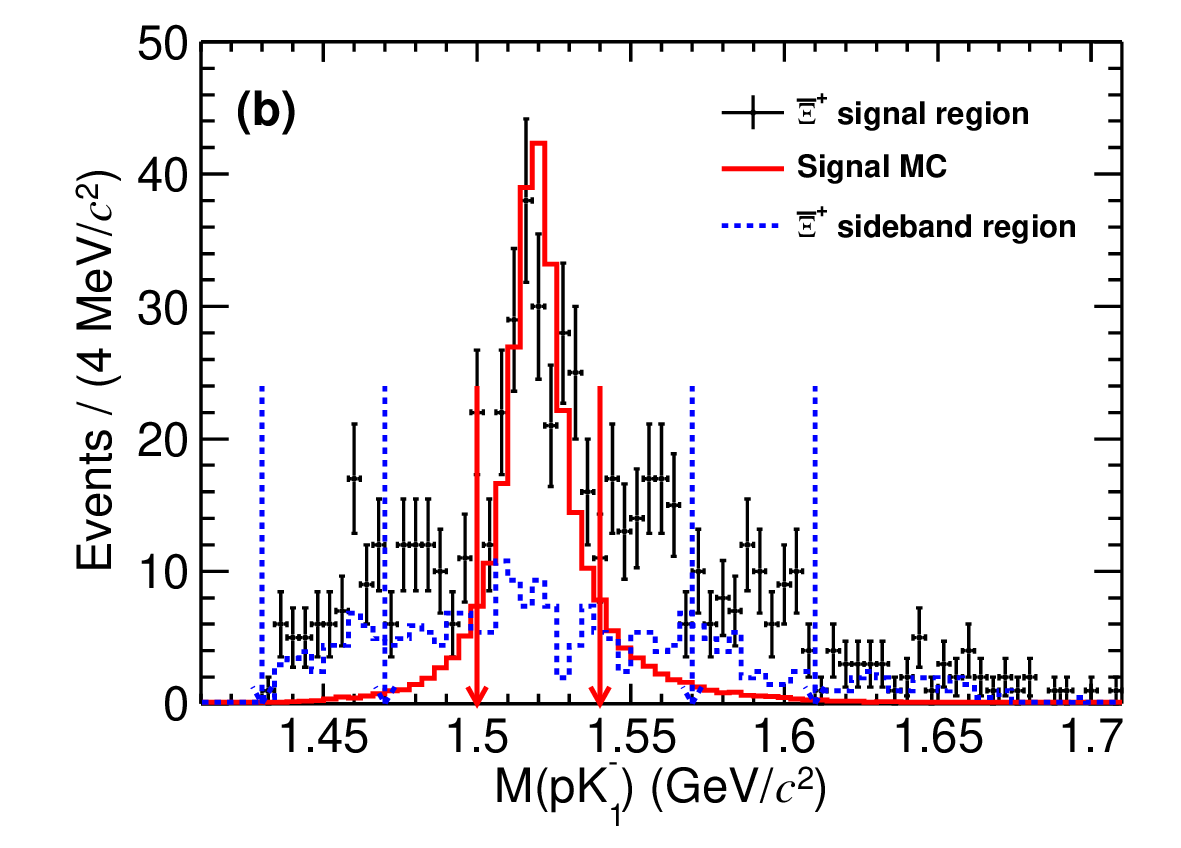}
\end{minipage}
\caption{(a) The distribution of $RM(pK^{-}_{1}K^{-}_{2})$ from data. The red solid and purple dashed arrows denote the $\bar{\Xi}^+$ signal and sideband regions, respectively; (b) The distribution of $M(pK^{-}_{1})$ in the $\bar{\Xi}^+$ signal and sideband regions from both data and MC simulation, where the normalization factor has been applied to the sideband distribution. The red solid and blue dashed arrows denote the $\Lambda(1520)$ signal and sideband regions, respectively.}
\label{fig:sideband1}
\end{center}
\end{figure}

Figure~\ref{fig:2dsideband} shows the 2D distribution of the recoil mass of $pK^{-}_{1}K^{-}_{2}$ versus the mass of $pK^{-}_{1}$ for data events, where the red box denotes the signal region of $\bar{\Xi}^+-\Lambda(1520)$, while the eight boxes with the same area around signal region are taken as the 2D sideband. Figure~\ref{fig:sideband1}(a) shows the $RM(pK^{-}_{1}K^{-}_{2})$ distribution of the survived candidate events. A clear $\bar{\Xi}^+$ signal is observed. We define the recoil mass range 1.31 GeV$/c^{2} <RM(pK^{-}_{1}K^{-}_{2})<$ 1.34 GeV$/c^{2}$ as the $\bar{\Xi}^+$ signal region, while the sideband regions are defined as 1.242 GeV$/c^{2} <RM(pK^{-}_{1}K^{-}_{2})<$ 1.272 GeV$/c^{2}$ and 1.372 GeV$/c^{2} <RM(pK^{-}_{1}K^{-}_{2})<$ 1.402 GeV$/c^{2}$. Figure~\ref{fig:sideband1}(b) shows the $M(pK^{-}_{1})$ distribution of the events in the $\bar{\Xi}^+$ signal and sideband regions, and a clear $\Lambda(1520)$ signal is observed, while no obvious peaking background is found in the $\bar{\Xi}^+$ sideband events. The $\Lambda(1520)$ signal region is defined as 1.50 GeV$/c^{2} <M(pK^{-}_{1})<$ 1.54 GeV$/c^{2}$, while the sideband regions are defined as 1.43 GeV$/c^{2} <M(pK^{-}_{1})<$ 1.47 GeV$/c^{2}$ and 1.57 GeV$/c^{2} <M(pK^{-}_{1})<$ 1.61 GeV$/c^{2}$. The normalization factor ($f_{\rm sideband}$) of the $\bar{\Xi}^+$ signal over the sideband regions is determined to be $0.47 \pm 0.01$ ($0.49 \pm 0.01$ for c.c. mode) according to the areas of the fitted background function between signal and sideband regions, as shown in Fig.~\ref{fig:sideband1}(a). This factor is then used to normalize the number of non-$\bar{\Xi}^+$ background events estimated from the sideband regions.

\section{Background study}
\label{Background}
\begin{figure}[htp]
\begin{center}
\begin{minipage}[t]{0.6\linewidth}
\includegraphics[width=1\textwidth]{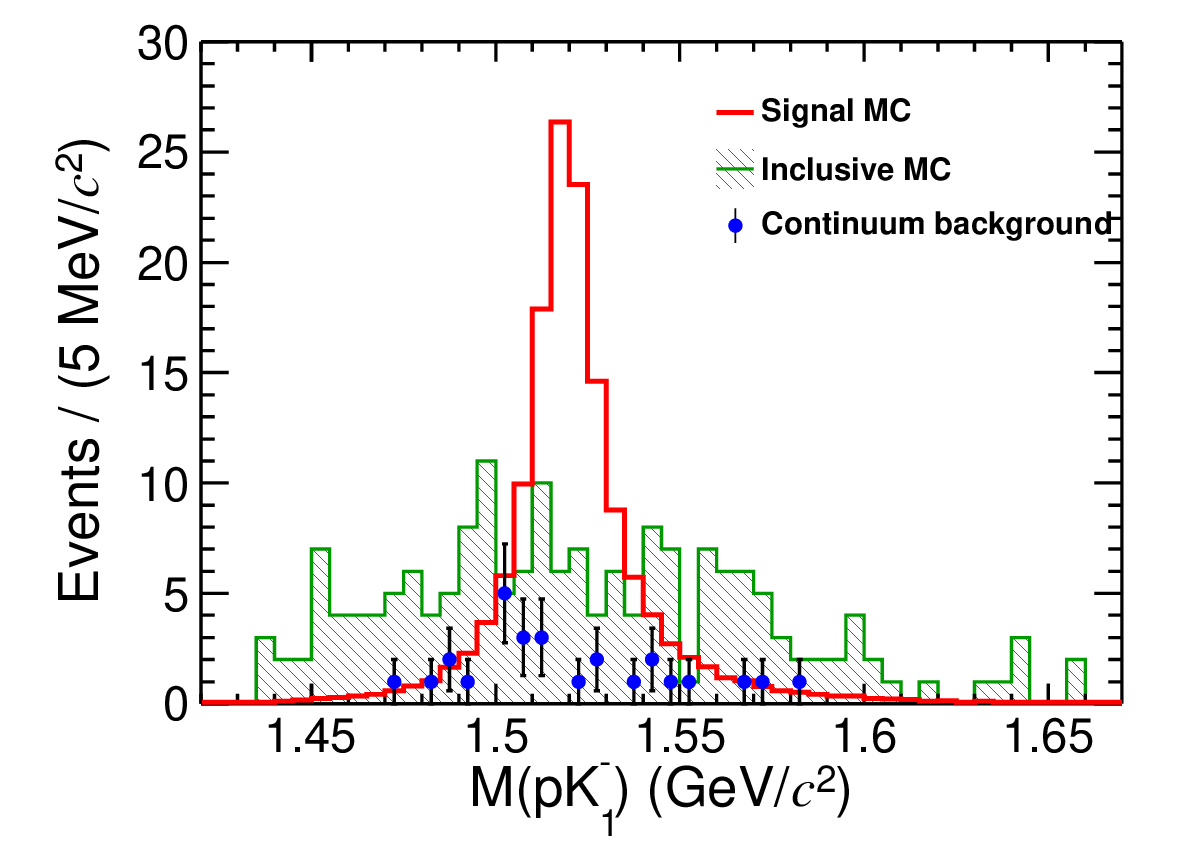}
\end{minipage}
\caption{The distributions of $M(pK^{-}_{1})$ from the inclusive MC sample and the continuum data. The red solid line denotes the signal MC. The green-shaded histogram denotes the inclusive MC. The dots with error bars represent the continuum background.}
\label{fig:after}
\end{center}
\end{figure}

The inclusive MC sample is used to investigate the potential backgrounds. Figure~\ref{fig:after} shows the distribution of the invariant mass of $pK_1$ for the accepted candidate events from simulated signal and background samples. Additionally,  the events selected from the sample collected at $\sqrt{s}$ = 3.773 GeV are analyzed to investigate the continuum production. No significant peaking background is found in either the inclusive MC sample or the continuum sample.

\section{Intermediate state}
\label{Intermediate}
\begin{figure}[htp]
\begin{center}
\begin{minipage}[t]{0.32\linewidth}
\includegraphics[width=1\textwidth]{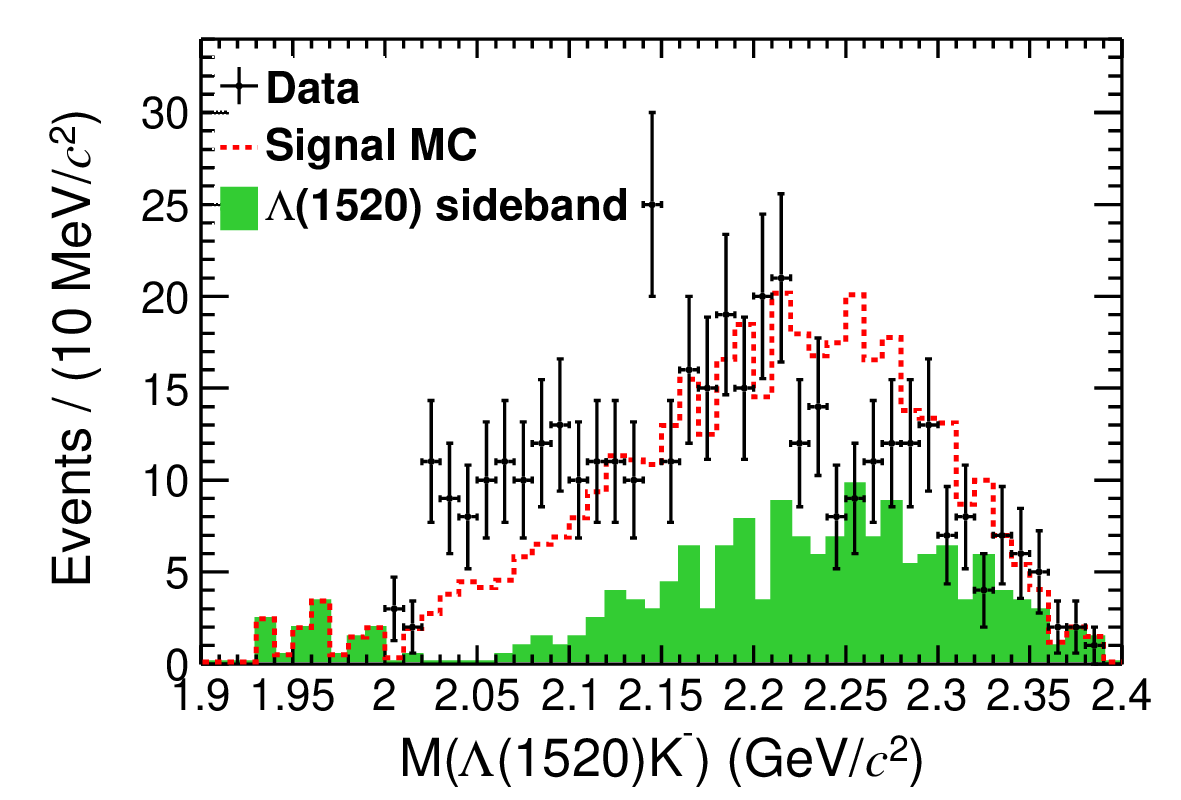}
\end{minipage}
\begin{minipage}[t]{0.32\linewidth}
\includegraphics[width=1\textwidth]{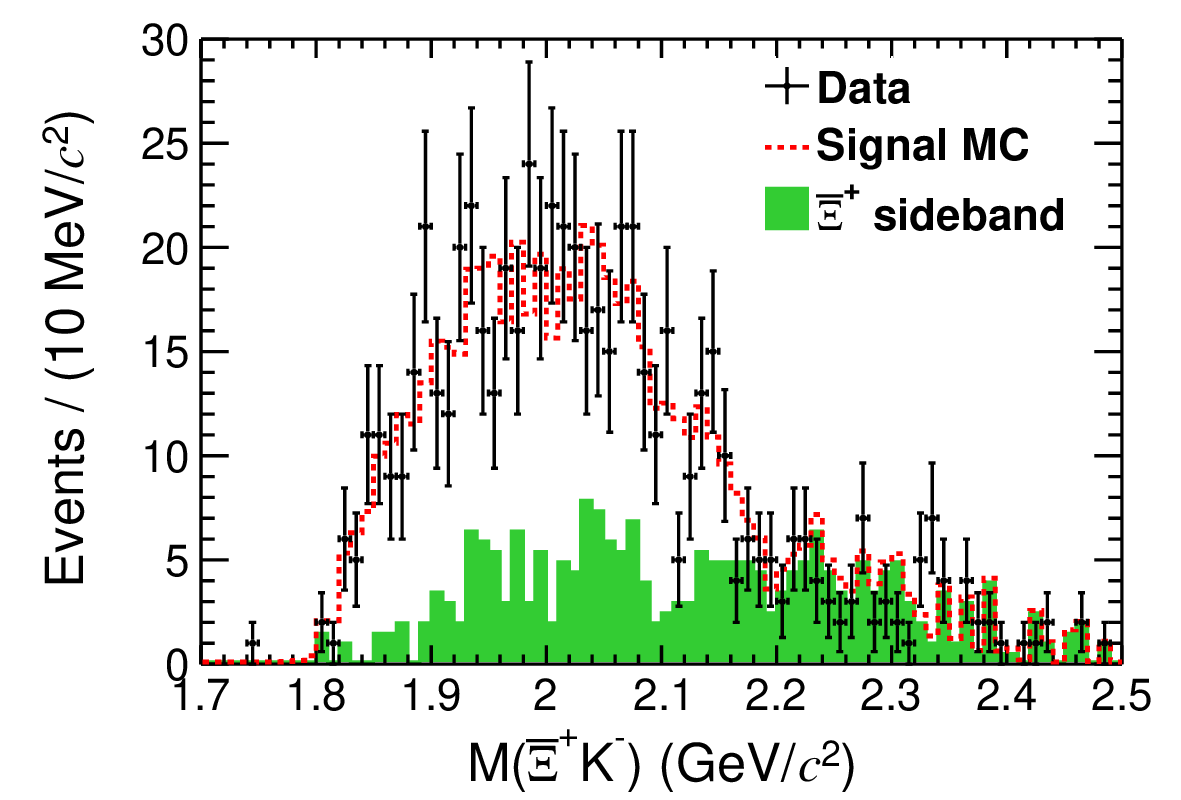}
\end{minipage}
\begin{minipage}[t]{0.32\linewidth}
\includegraphics[width=1\textwidth]{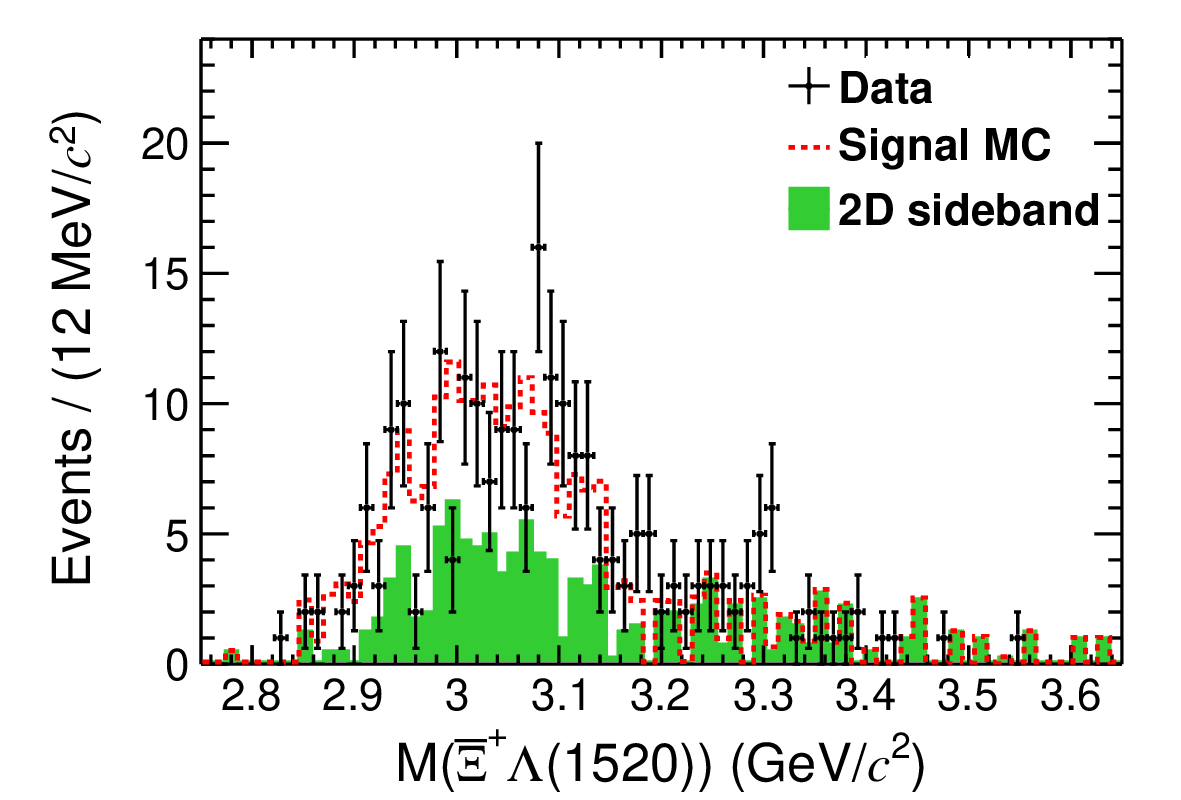}
\end{minipage}
\caption{(a,b,c) The distributions of $M(\Lambda(1520)K^{-})$, $M(\bar{\Xi}^{+}K^{-})$, and $M(\bar{\Xi}^{+}\Lambda(1520))$ for data and signal MC events.}
\label{fig:intermediate}
\end{center}
\end{figure}

To search for the potential intermediate states, we investigate the invariant mass spectra for all two-body combinations with the events in the $\bar{\Xi}^{+}$ signal region (1.31 GeV$/c^{2} <RM(pK^{-}_{1}K^{-}_{2})<$ 1.34 GeV$/c^{2}$) or $\Lambda(1520)$ signal region (1.50 GeV$/c^{2} <M(pK^{-}_{1})<$ 1.54 GeV$/c^{2}$). Figure~\ref{fig:intermediate} shows the distributions of $M(\Lambda(1520)K^{-})$, $M(\bar{\Xi}^{+}K^{-})$, and $M(\bar{\Xi}^{+}\Lambda(1520))$ for data and MC events generated with PHSP model, where the contributions from the normalized sideband have been added to MC sample. No obvious intermediate structure is observed in each distribution.

\section{Signal yield and branching fraction}
In investigating the intermediate states, we require one entry per event, which is likely to distort the background shape. To avoid this issue, the signal yield for $\psi(3686) \to K^{-}\Lambda(1520)\bar{\Xi}^{+}$, $\Lambda(1520) \to pK^{-}$, is obtained by performing an unbinned maximum likelihood fit to the combined $M(pK^{-}_{1})$ and $M(pK^{-}_{2})$ distributions with dual entries per event. The signal shape is described by the signal MC shape convolved with a Gaussian function which accounts for the difference in mass and mass resolution between data and MC simulation. The parameters of this Gaussian function are free to float in the fit.

The background components consist of the non-$\bar{\Xi}^{+}$,
non-$\Lambda(1520)$, and continuum backgrounds. The shape and yield of
the non-$\bar{\Xi}^{+}$ background is fixed in the fit according to
the $\bar{\Xi}^{+}$ data sidebands after smoothing and
normalization. The non-$\Lambda(1520)$ background from the four-body
process ($\psi(3686) \to pK^{-}K^{-}\bar{\Xi}^{+}$) is described by
the MC-simulated shape with its yield floating in the fit.  The shape
of the continuum background is taken from the data distribution at
$\sqrt{s}=3.773$ GeV as described in Sec.\ref{Background}. To account
for the difference of the production cross sections and integrated
luminosities between the two energies, the continuum background
yield is scaled by a factor $f_{c}$ calculated as
\begin{equation}\label{con}
\textstyle{\displaystyle f_{c} = \frac{\mathcal{L}_{3.686}}{\mathcal{L}_{3.773}} \times \frac{\sigma_{3.686}}{\sigma_{3.773}}},
\end{equation} where $\mathcal{L}_{3.686} = 3877.05 pb^{-1}$ and
$\mathcal{L}_{3.773} = 2931.8 pb^{-1}$ are the integrated luminosities
at 3.686 GeV and 3.773 GeV, respectively; $\sigma$ denotes the production
cross section, which is assumed to be proportional to $1/s$. The value
of $f_{c}$ is determined to be 1.386.

\begin{figure}[htb]
\begin{center}
\begin{minipage}[t]{0.6\linewidth}
\includegraphics[width=1\textwidth]{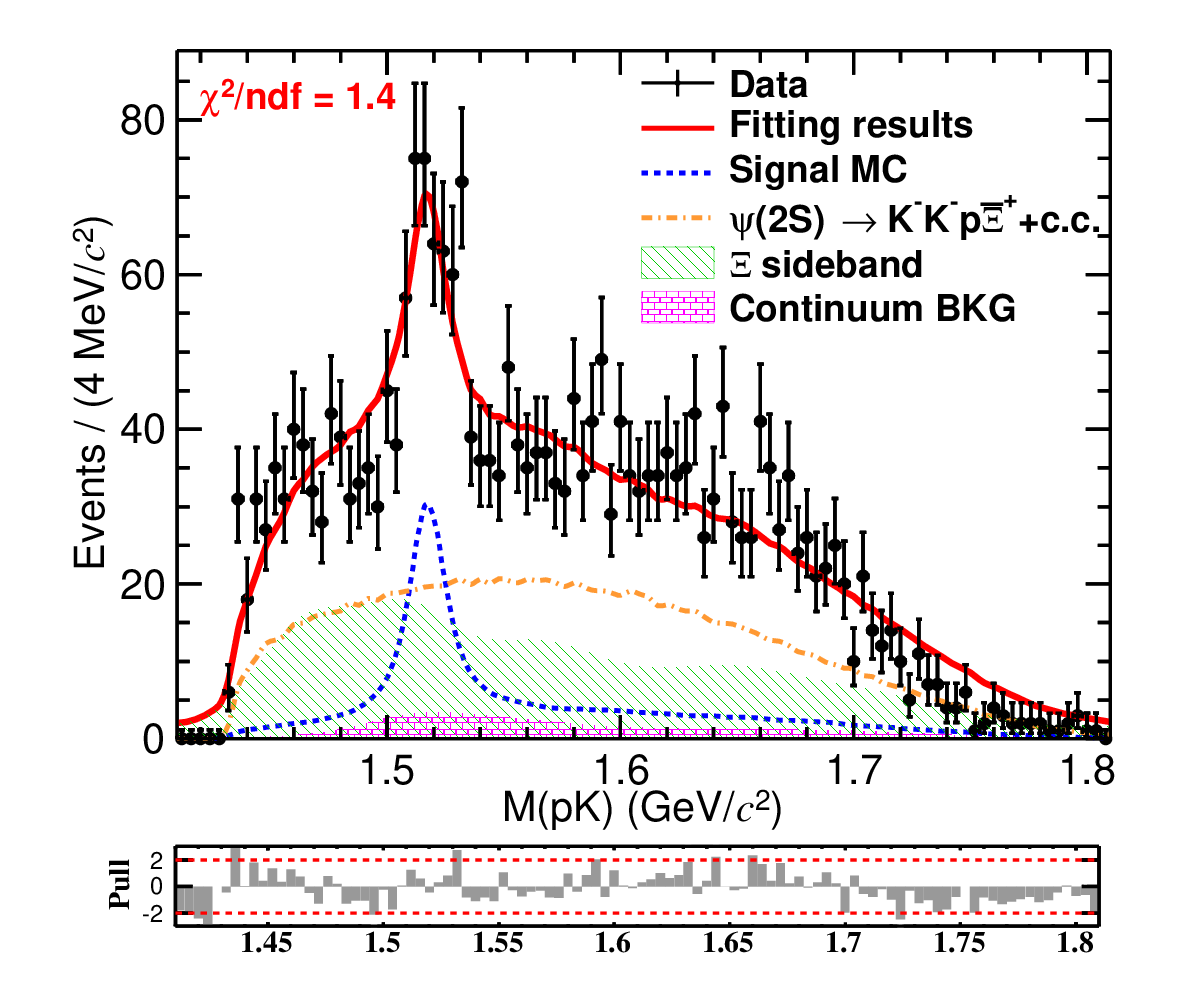}
\end{minipage}
\caption{Fit to the $M_{pK}$ distribution of the two charge conjugate
  signal modes. The dots with error bars represent the data sample
  events. The red solid line represents the fitting results. The blue
  dashed line represents the signal. The orange, green, and purple dashed lines denote the backgrounds from the non-$\Lambda(1520)$, non-${\Xi}$, and continuum processes, respectively.}
\label{fig:fit1}
\end{center}
\end{figure}

We fit to the dataset of the two charge conjugate channels together, which is taken as the nominal result. The goodness of the fit is $\chi^{2}/ndf$ = 108/76, and the statistical significance of signal is 6.9$\sigma$. The significance is determined by comparing the likelihood difference with or without including $\Lambda(1520)$ signal contribution, and taking into account the change in the number of degrees of freedom~\cite{wilks1938large}. Figure~\ref{fig:fit1} shows the fit result for the sum of two charge conjugate channels, and the fitted number of $\Lambda(1520)$ signal is 190 $\pm$ $15_{\rm{stat}.}$, where the uncertainty is statistical. The nominal detection efficiency for the $\psi(3686) \to K^{-}\Lambda(1520)\bar{\Xi}^{+} + c.c.$ channel is 11.55$\%$, which is obtained as the average of the detection efficiencies of the two charge conjugate channels.

The product branching fraction ${\cal B}[\psi(3686) \to K^{-}\Lambda(1520)\bar{\Xi}^{+} + c.c.] \times {\cal B}[\Lambda(1520) \to pK^{-}]$ is calculated via

\begin{footnotesize}
\begin{equation}\label{bf}
\textstyle{\displaystyle\mathcal{B}[\psi(3686) \to K^{-}\Lambda(1520)\bar{\Xi}^{+} + c.c.] \times \mathcal{B}[\Lambda(1520) \to pK^{-}] = \frac{N^{\rm obs}}{N_{\psi(3686)}\mathcal{B}(\bar{\Xi}^{+} \to \bar{\Lambda}\pi^{+})\mathcal{B}(\bar{\Lambda} \to \bar{p}\pi^{+})\epsilon}} ,
\end{equation}
\end{footnotesize}where $N^{\rm obs}$ represents the number of observed signal events; $N_{\psi(3686)}$ represents the total number of $\psi(3686)$ events in data; $\displaystyle\mathcal{B}(\bar{\Xi}^{+} \to \bar{\Lambda}\pi^{+})$ and $\displaystyle\mathcal{B}(\bar{\Lambda} \to \bar{p}\pi^{+})$ denote the branching fractions of $\bar{\Xi}^{+} \to \bar{\Lambda}\pi^{+}$ and $\bar{\Lambda} \to \bar{p}\pi^{+}$, respectively, which are taken from PDG~\cite{ParticleDataGroup:2024cfk}; $\epsilon$ is the detection efficiency. Finally, the product branching fraction ${\cal B}[\psi(3686) \to K^{-}\Lambda(1520)\bar{\Xi}^{+} + c.c.] \times {\cal B}[\Lambda(1520) \to pK^{-}]$ is measured to be $(9.5 \pm 0.8) \times 10^{-7}$, where the uncertainty is statistical only.

\section{Systematic uncertainty}
\label{Sys}

The systematic uncertainties in the branching fraction measurement are from the tracking efficiency, PID efficiency, signal shape, non-$\Lambda(1520)$ background, continuum background, $\Xi$ mass window, $\Xi$ sideband, MC imperfection, MC statistics, total number of $\psi(3686)$ events, and the cited branching fractions. The details are discussed below:

\begin{itemize}

    \item Tracking efficiency. Based on the control samples $J/\psi \to K^{0}_{S}K^{\pm}\pi^{\mp}$, and $J/\psi \to p\Bar{p}\pi^{+}\pi^{-}$, the relative difference of the MDC tracking efficiencies between data and MC simulation has been estimated in the polar angle and momentum distributions. The uncertainties related to tracking for kaons, pions, protons, and antiprotons are estimated to be 0.7$\%$, 1.6$\%$, 0.6$\%$, and 0.5$\%$, respectively.

    \item PID efficiency. Based on the control samples $J/\psi \to K^{0}_{S}K^{\pm}\pi^{\mp}$, and $J/\psi \to p\Bar{p}\pi^{+}\pi^{-}$, the relative difference of the PID efficiencies between data and MC simulation has been estimated in the polar angle and momentum distributions. The uncertainties related to PID for kaons, pions, protons, and antiprotons are estimated to be 0.1$\%$, 1.0$\%$, 0.3$\%$, 0.5$\%$, respectively.

    \item Signal shape. The uncertainty related to the signal shape is
      estimated by replacing the MC-simulated shape with an
      alternative one, in which a Breit-Wigner function convolved with
      a double Gaussian is used to model the $\Lambda(1520)$ and the
      MC simulated shape is utilized to describe the wrong $pK^{-}$
      combination, with the yield ratio of these two
      components fixed to one. The difference of the fitted signal yield with
      the nominal one, 5.0$\%$, is assigned as the uncertainty.

    \item Non-$\Lambda(1520)$ background.
    The uncertainty due to the non-$\Lambda(1520)$ background is estimated by replacing the background shape from the MC-simulated shape to a seventh-order polynomial multiplied by an Argus function~\cite{ARGUS:1990hfq}. The difference of resulted signal yield with the nominal one is taken as the uncertainty, which is 4.1$\%$.

    \item Continuum background. The uncertainty due to the event number of continuum background is estimated to be 1.0$\%$, by varying the continuum background yield in the fit by one standard deviation, assuming it follows a Poisson distribution. The uncertainty caused by the continuum background shape is estimated to be 5.0$\%$, by replacing the background shape from RooKeysPdf~\cite{Cranmer:2000du} to RooHistPdf~\cite{Antcheva:2009zz}. By combining them together, the uncertainty related to the continuum background is taken as 5.1$\%$.

    \item $\Xi$ mass window. The uncertainty due to the requirement of $\Xi$ mass window is estimated with the control sample $J/\psi \to \Xi^{-}\bar{\Xi}^{+}$. The relative efficiency of this requirement is calculated as $\frac{N_{with}}{N_{without}}$, where $N_{with}$ and $N_{without}$ denote the number of events obtained with and without the $\Xi$ mass window requirement. The difference of this efficiency between data and MC, 5.1$\%$, is taken as  the corresponding uncertainty.

    \item $\Xi$ sideband. The uncertainty caused by $\Xi$ sideband is estimated by varying the $\Xi$ sideband region from $[m_{\Xi}-0.08, m_{\Xi}-0.05] \cup [m_{\Xi}+0.05, m_{\Xi}+0.08]$ to $[m_{\Xi}-0.085, m_{\Xi}-0.055] \cup [m_{\Xi}+0.055, m_{\Xi}+0.085]$ GeV/$c^{2}$. The resulted uncertainty is negligible.

    \item MC imperfection. We check the cos$\theta$ and momentum distributions of pions, kaons, protons, and antiprotons from both data and signal MC samples, which are shown in Fig.~\ref{cormom}. There are some differences in the momentum distributions. To estimate the uncertainty caused by MC imperfection, the efficiency is weighted according to the data momentum distributions where the normalized backgrounds from the $\Xi$ sideband have been subtracted. The weighted efficiency is calculated via
\begin{equation}\label{corr}
\textstyle{\displaystyle \bar{\epsilon} = \frac{\sum_{i,j,k}n^{\rm MC}_{i,j,k} \cdot \epsilon_{i,j,k}}{\sum_{i,j,k}n^{\rm truth}_{i,j,k} \cdot \epsilon_{i,j,k}}} ,
\end{equation}
where
\begin{equation}\label{ratio}
\textstyle{ {\epsilon_{i,j,k}} =
\left\{
\begin{aligned}
&(n^{\rm data}_{i,j,k}-n^{\rm sideband}_{i,j,k} \cdot f_{\rm sideband})/n^{\rm MC}_{i,j,k} &,  & & n^{\rm MC}_{i,j,k} \not= 0  \\
&1 &, & & n^{\rm MC}_{i,j,k} = 0
\end{aligned}
\right.}.
\end{equation}Here, $n^{\rm MC}_{i,j,k}$, $n^{\rm truth}_{i,j,k}$, $n^{\rm data}_{i,j,k}$, and $n^{\rm sideband}_{i,j,k}$ refer to the numbers of events in the $(i,j,k)$-th bin of MC, MC truth, data, and $\Xi$ sideband samples. The associated uncertainty is determined to be 4.1$\%$.

\begin{figure}[htb]
\begin{center}
\includegraphics[width=.32\textwidth]{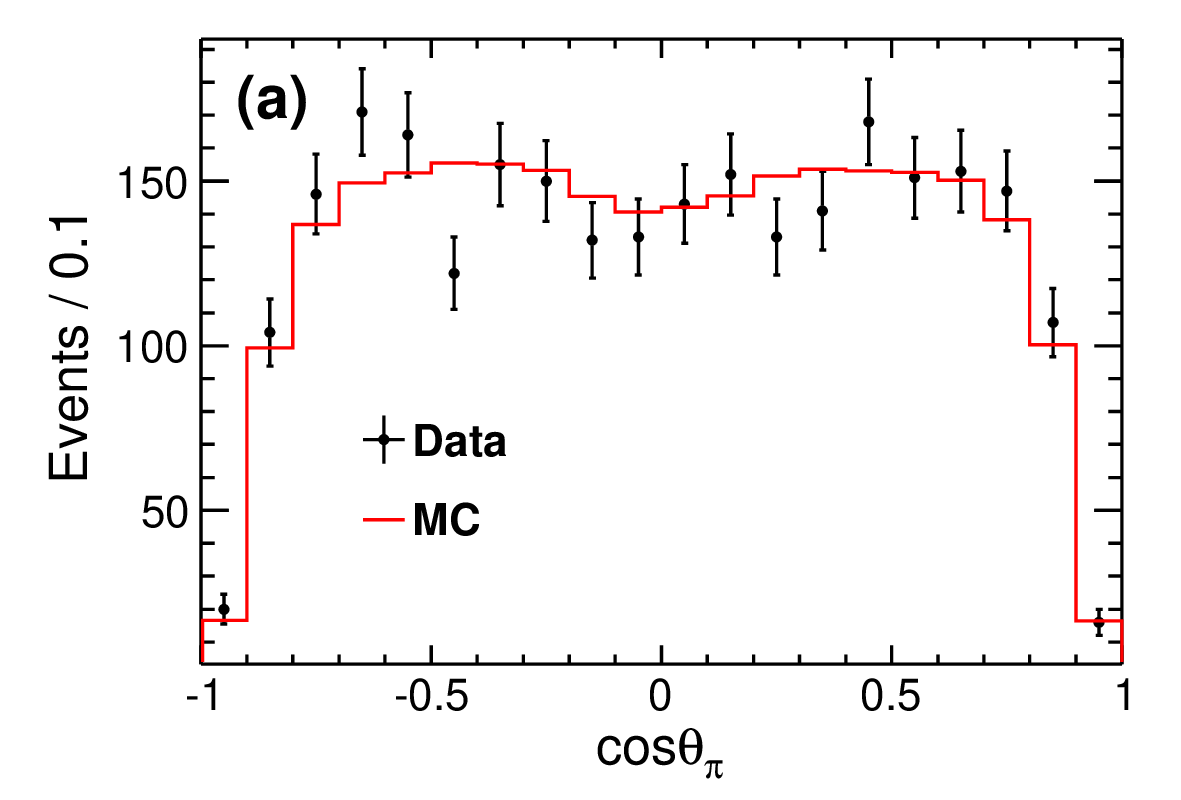}
\includegraphics[width=.32\textwidth]{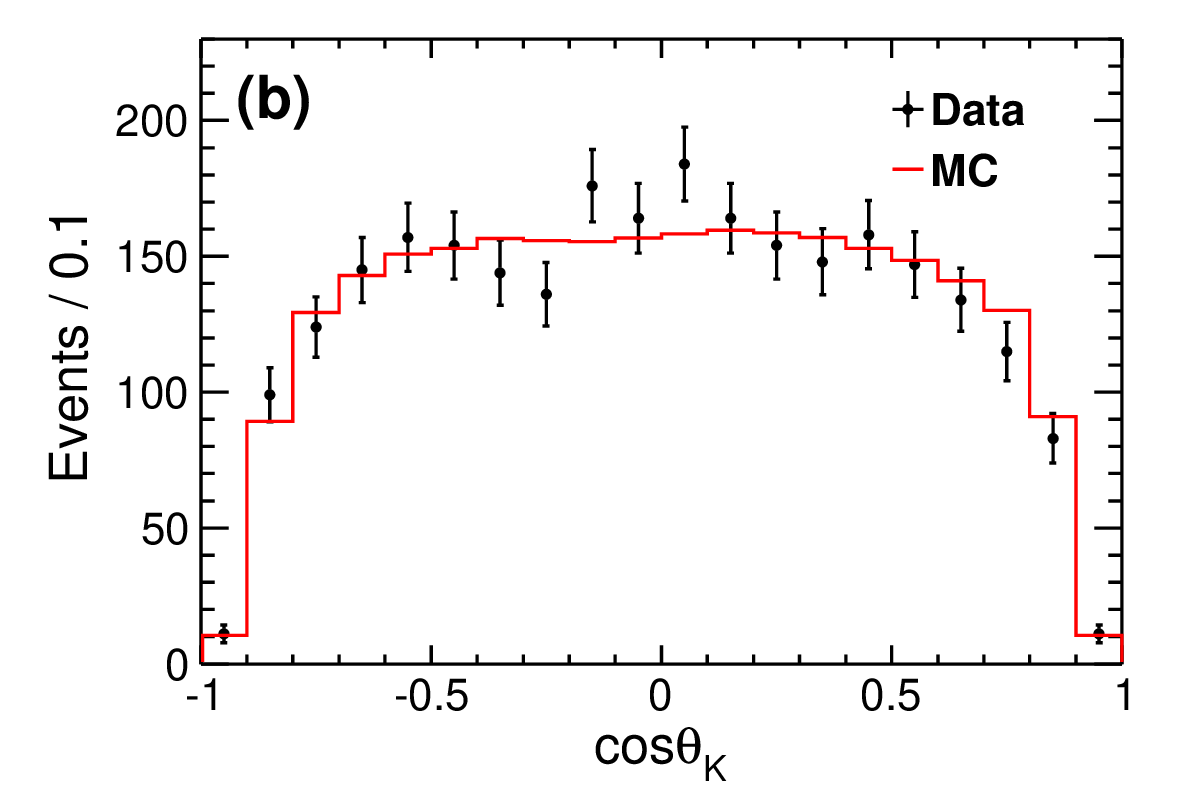}
\includegraphics[width=.32\textwidth]{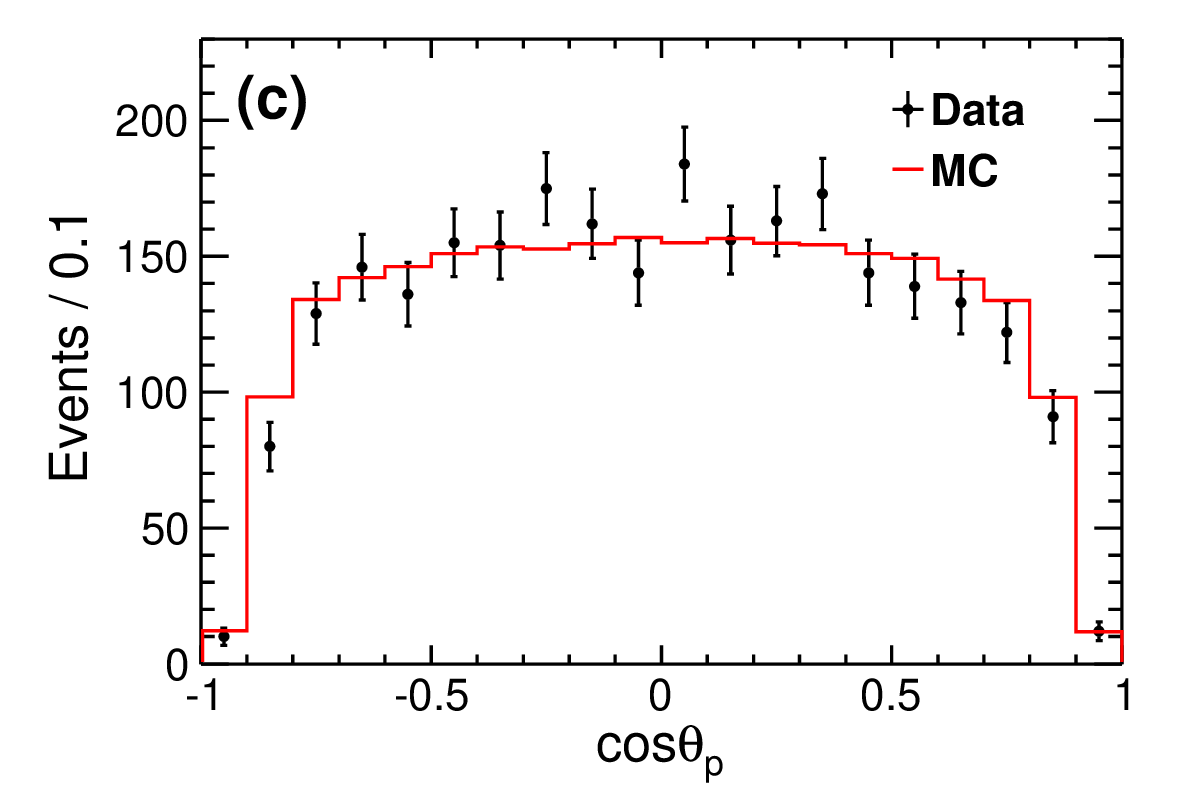}
\includegraphics[width=.32\textwidth]{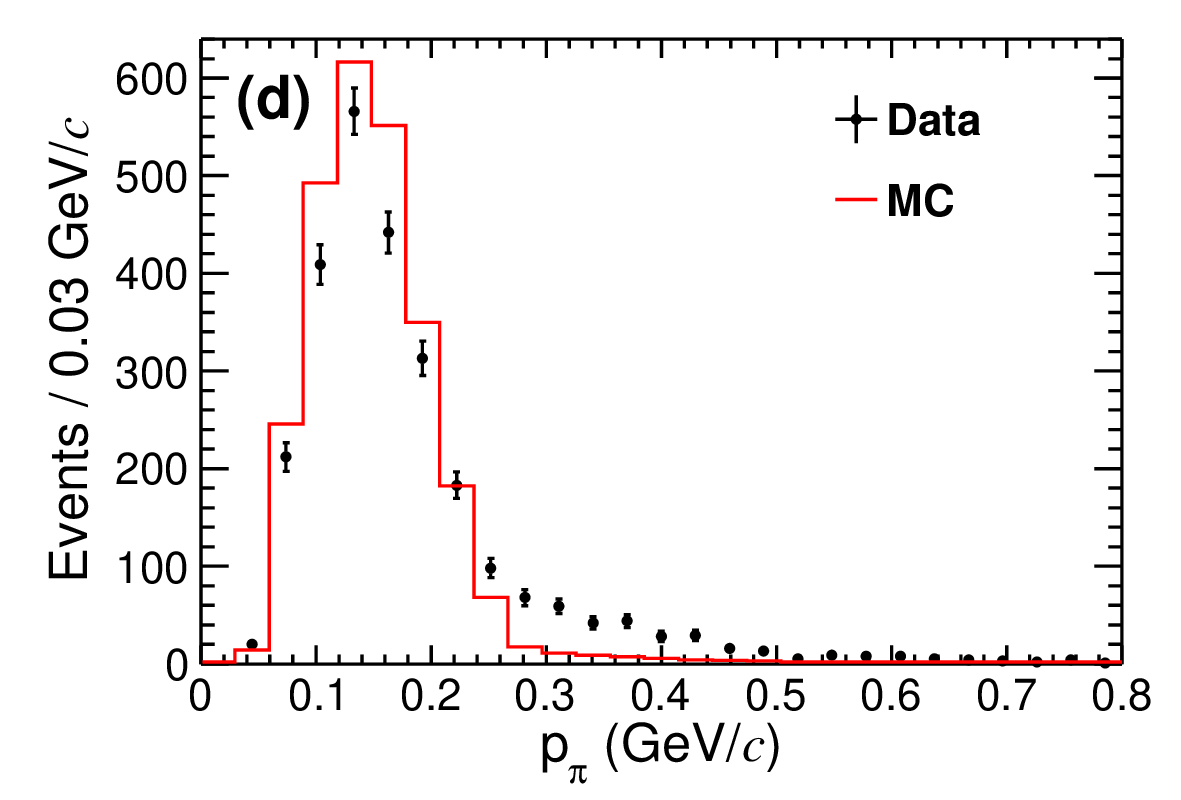}
\includegraphics[width=.32\textwidth]{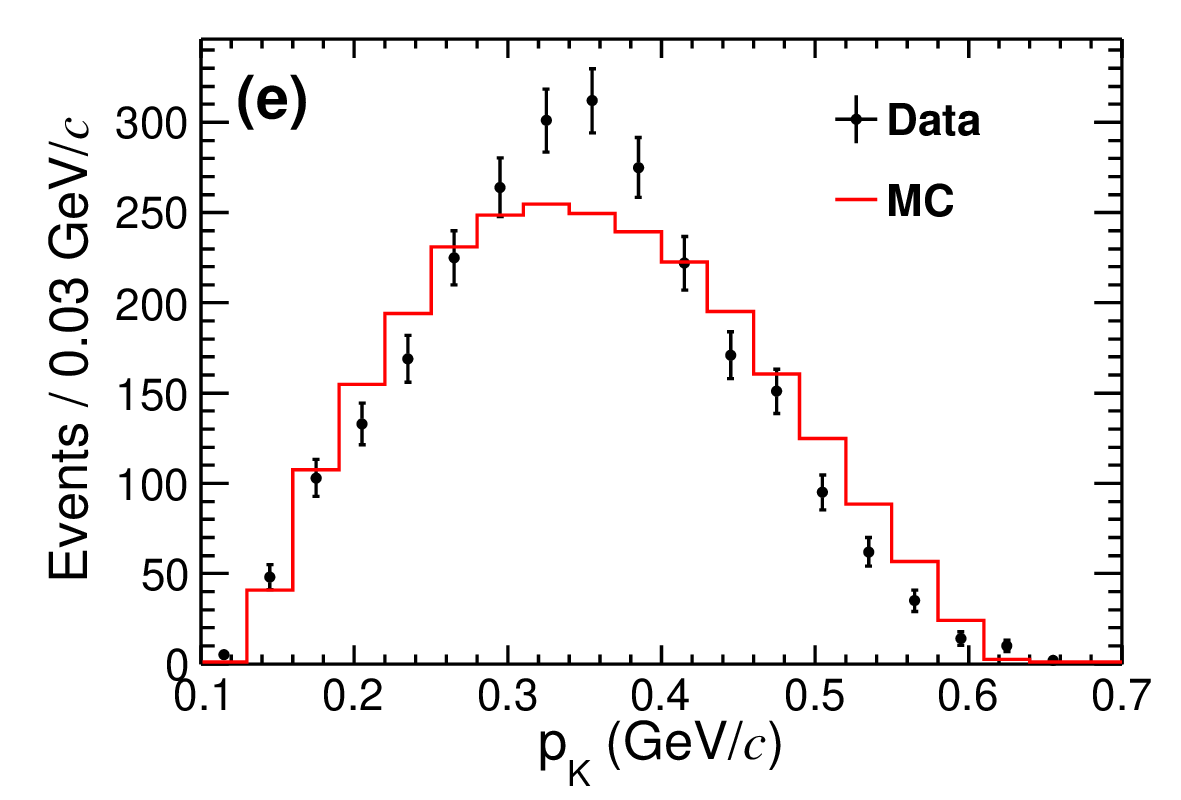}
\includegraphics[width=.32\textwidth]{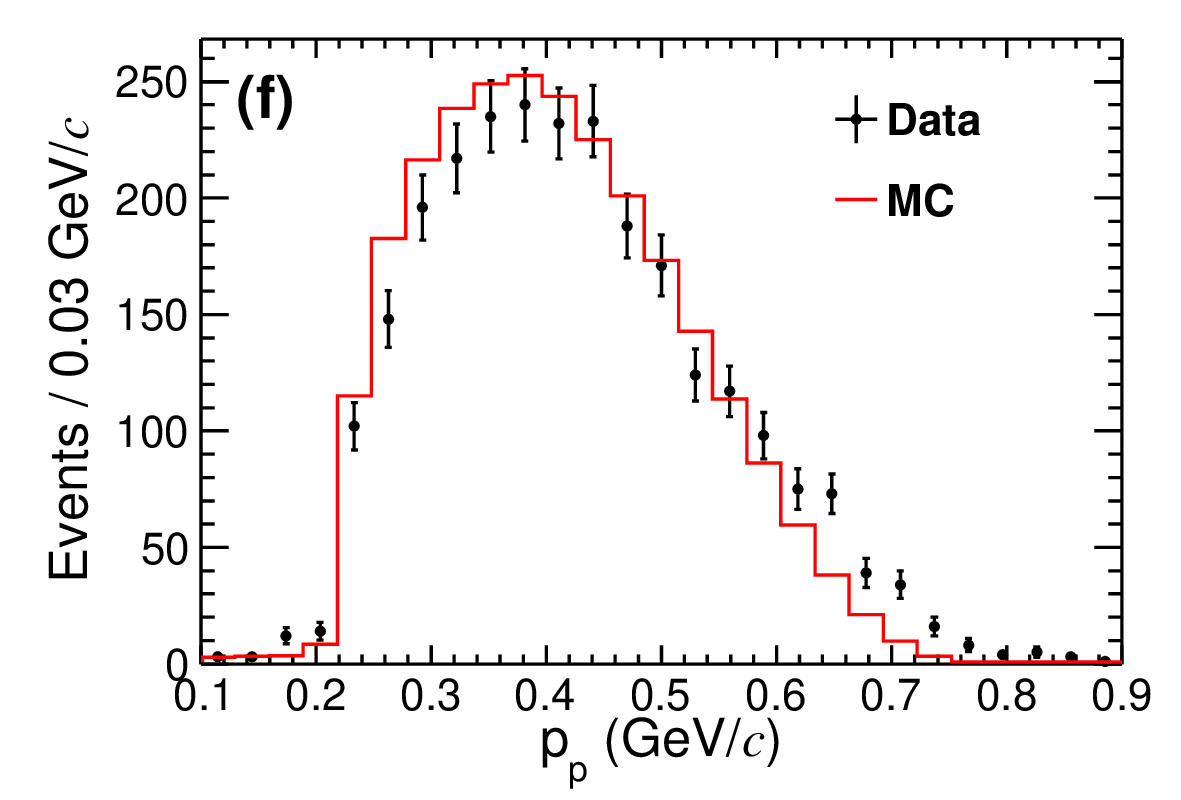}
\includegraphics[width=.32\textwidth]{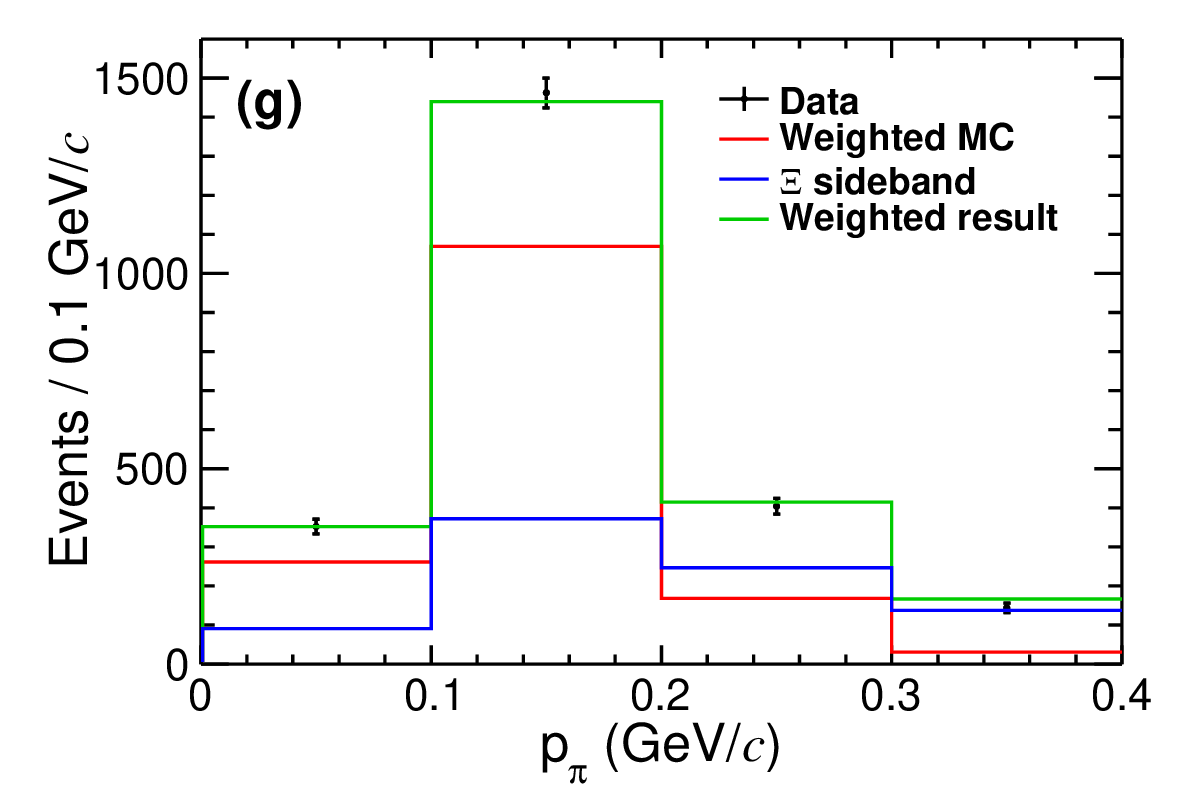}
\includegraphics[width=.32\textwidth]{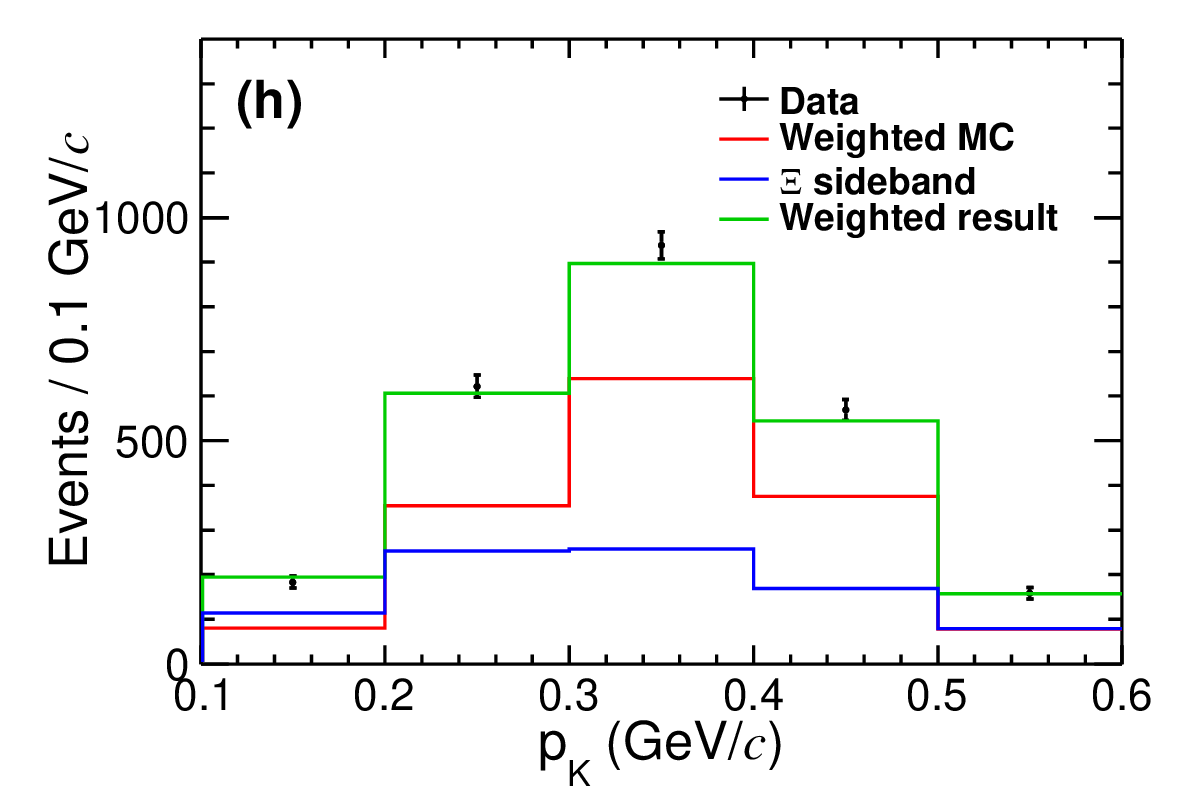}
\includegraphics[width=.32\textwidth]{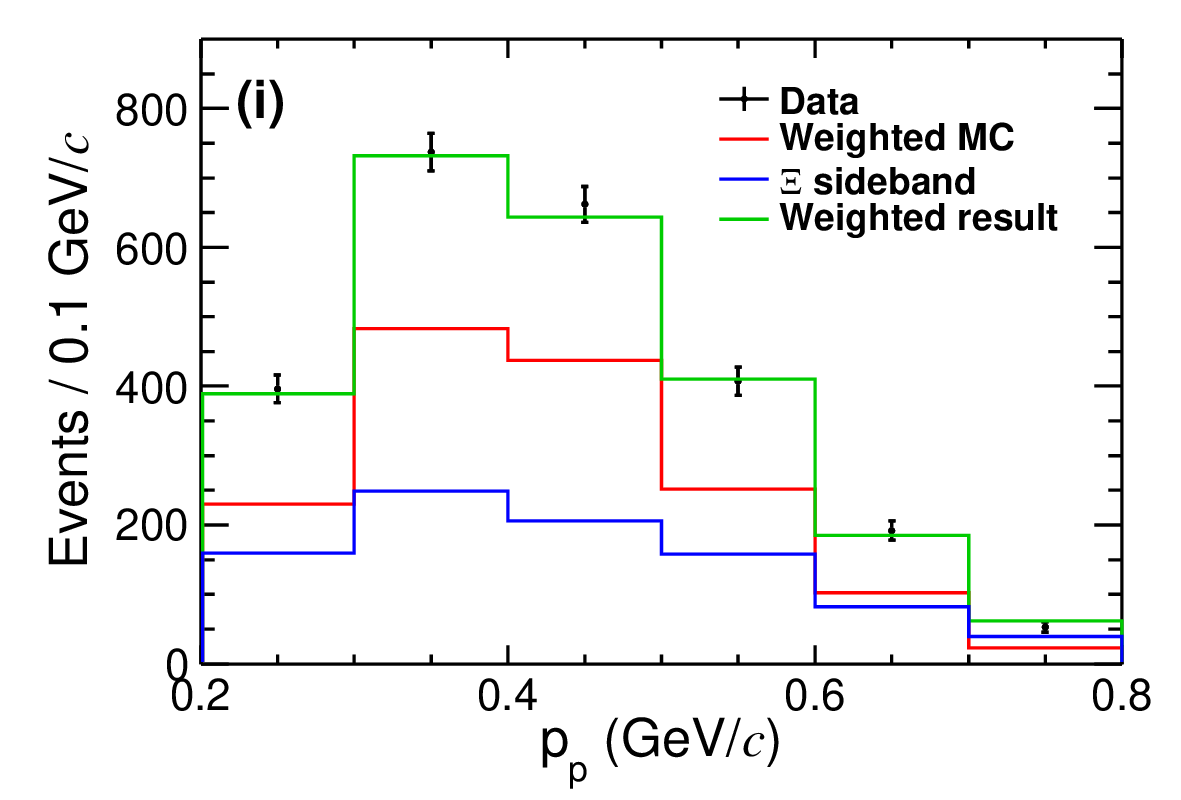}
\caption{The one-dimensional comparisons of cos$\theta$~(a,b,c), and
  momentum~(d,e,f) distributions of pions, kaons, protons, and
  antiprotons from data and signal MC samples. Projections of the
  weighted result based on the three-dimensional distribution of
  momentum is also shown~(g,h,i), where the dots with error bars
  represent the data; the red, and blue solid lines denote the
  weighted signal MC, and $\Xi$ sideband respectively and the green
  solid line is the sum of the two.  }
\label{cormom}
\end{center}
\end{figure}

    \item MC statistics. The uncertainty is negligible due to the large generated MC sample.

    \item Total number of $\psi(3686)$ events. The uncertainty of the total number of $\psi(3686)$ events is  0.5$\%$~\cite{BESIII:2024lks}.

    \item The quoted branching fractions. The uncertainty due to the quoted branching fraction $\Lambda \to p\pi^{+}$, is $0.8\%$ according to the PDG~\cite{ParticleDataGroup:2024cfk}. The uncertainty from the branching fraction of $\Xi^{-} \to \Lambda\pi^{-}$ is negligible.
\end{itemize}

All of the systematic uncertainties are summarized in Table~\ref{BF-Sys}. Adding them in quadrature results in a total systematic uncertainty of 11.3$\%$ in the branching fraction measurement.

\begin{table}[htbp]
    \caption{Systematic uncertainties in the branching fraction measurement.}
  \label{BF-Sys}
  \centering
  \begin{tabular}{lc}
  \hline
    Source   & Uncertainty~(\%) \\
  \hline
    Tracking  & 3.4\\
    PID  & 1.9\\
    Signal shape     & 5.0\\
    Non-$\Lambda(1520)$ background & 4.1\\
    Continuum background & 5.1\\
$\Xi$ mass window  & 5.1\\
    $\Xi$ sideband & Negligible\\
    MC imperfection & 4.1\\
    MC statistics & Negligible\\
    Total number of $\psi(3686)$ events &  0.5\\
    Quoted branching fractions &  0.8\\
    \hline
    Total            & 11.3\\
    \hline
  \end{tabular}
\end{table}

Additionally, the signal significance is updated to be 5.0$\sigma$ after considering the uncertainties related to the signal shape, non-$\Lambda(1520)$ background, continuum background, and $\Xi$ mass window.

\section{Summary}
In summary, using a sample of 2.7 billion $\psi(3686)$ events
collected with the BESIII detector, the decay $\psi(3686) \to
K^{-}\Lambda(1520)\bar{\Xi}^{+} + c.c.$ is observed for the first time
with a significance of 5.0 standard deviations, after considering both statistical and systematic uncertainties. The product branching fraction ${\cal B}[\psi(3686) \to K^{-}\Lambda(1520)\bar{\Xi}^{+} + c.c.] \times {\cal B}[\Lambda(1520) \to pK^{-}]$ is measured to be $(9.5 \pm 0.8 \pm 1.1) \times 10^{-7}$, where the first uncertainty is statistical and the second systematic. With the current sample size, no evidence of excited baryon state or threshold enhancement is observed. A larger $\psi(3686)$ data sample may help to study the dynamics of this decay.

\acknowledgments
The BESIII Collaboration thanks the staff of BEPCII and the IHEP computing center for their strong support. This work is supported in part by National Key R\&D Program of China under Contracts Nos. 2020YFA0406300, 2020YFA0406400, 2023YFA1606000; National Natural Science Foundation of China (NSFC) under Contracts Nos. 11635010, 11735014, 11935015, 11935016, 11935018, 12025502, 12035009, 12035013, 12061131003, 12192260, 12192261, 12192262, 12192263, 12192264, 12192265, 12221005, 12225509, 12235017, 12361141819; the Chinese Academy of Sciences (CAS) Large-Scale Scientific Facility Program; the CAS Center for Excellence in Particle Physics (CCEPP); Joint Large-Scale Scientific Facility Funds of the NSFC and CAS under Contract No. U1832207; 100 Talents Program of CAS; The Institute of Nuclear and Particle Physics (INPAC) and Shanghai Key Laboratory for Particle Physics and Cosmology; German Research Foundation DFG under Contracts Nos. FOR5327, GRK 2149; Istituto Nazionale di Fisica Nucleare, Italy; Knut and Alice Wallenberg Foundation under Contracts Nos. 2021.0174, 2021.0299; Ministry of Development of Turkey under Contract No. DPT2006K-120470; National Research Foundation of Korea under Contract No. NRF-2022R1A2C1092335; National Science and Technology fund of Mongolia; National Science Research and Innovation Fund (NSRF) via the Program Management Unit for Human Resources \& Institutional Development, Research and Innovation of Thailand under Contracts Nos. B16F640076, B50G670107; Polish National Science Centre under Contract No. 2019/35/O/ST2/02907; Swedish Research Council under Contract No. 2019.04595; The Swedish Foundation for International Cooperation in Research and Higher Education under Contract No. CH2018-7756; U. S. Department of Energy under Contract No. DE-FG02-05ER41374.

\bibliographystyle{JHEP}
\normalem
\bibliography{references}

\clearpage
\appendix
\clearpage{}\large
The BESIII Collaboration\\
\normalsize
M.~Ablikim$^{1}$, M.~N.~Achasov$^{4,c}$, P.~Adlarson$^{76}$, O.~Afedulidis$^{3}$, X.~C.~Ai$^{81}$, R.~Aliberti$^{35}$, A.~Amoroso$^{75A,75C}$, Q.~An$^{72,58,a}$, Y.~Bai$^{57}$, O.~Bakina$^{36}$
% ...
\\
\vspace{0.2cm} {\it
$^{1}$ Institute of High Energy Physics, Beijing 100049, People's Republic of China\\
% ...
$^{3}$ Bochum  Ruhr-University, D-44780 Bochum, Germany\\
$^{4}$ Budker Institute of Nuclear Physics SB RAS (BINP), Novosibirsk 630090, Russia\\
% ...
$^{35}$ Johannes Gutenberg University of Mainz, Johann-Joachim-Becher-Weg 45, D-55099 Mainz, Germany\\
$^{36}$ Joint Institute for Nuclear Research, 141980 Dubna, Moscow region, Russia\\
% ...
$^{57}$ Southeast University, Nanjing 211100, People's Republic of China\\
$^{58}$ State Key Laboratory of Particle Detection and Electronics, Beijing 100049, Hefei 230026, People's Republic of China\\
% ...
$^{72}$ University of Science and Technology of China, Hefei 230026, People's Republic of China\\
% ...
$^{75}$ University of Turin and INFN, (A)University of Turin, I-10125, Turin, Italy; (B)University of Eastern Piedmont, I-15121, Alessandria, Italy; (C)INFN, I-10125, Turin, Italy\\
$^{76}$ Uppsala University, Box 516, SE-75120 Uppsala, Sweden\\
% ...
$^{81}$ Zhengzhou University, Zhengzhou 450001, People's Republic of China\\

\vspace{0.2cm}
$^{a}$ Deceased\\
% ...
$^{c}$ Also at the Novosibirsk State University, Novosibirsk, 630090, Russia\\
% ...
}
\clearpage{}

\end{document}